\documentstyle[aps,psfig,prl,multicol]{revtex}

\begin{document}

\draft
\title{Filling a Silo with a Mixture of Grains:  \\ 
Friction-Induced Segregation} 

\author{Antal K\'arolyi$^{1-3}$, J\'anos Kert\'esz$^1$, Shlomo
Havlin$^{3,4}$, \\Hern\'an A. Makse$^{3,5}$ and H. Eugene Stanley$^3$}

\address{$^1$ Department of Theoretical Physics, Technical Univ. of 
Budapest, Budafoki u. 8, 1111 Hungary\\
$^2$ Theoretische Physik, FB 10, Gerhard-Mercator Universit\"at, 47048
Duisburg, Germany\\
$^3$ Center for Polymer Studies and Physics Dept., Boston University,
Boston, MA 02215 \\
$^4$Minerva Center and 
Department of Physics, Bar-Ilan University, Ramat Gan, Israel\\
$^5$ Schlumberger-Doll Research, Old Quarry Road, Ridgefield, CT 06877}

\date{\today}

\maketitle

\begin{abstract}

We study the filling process of a two-dimensional silo with inelastic
particles by simulation of a granular media lattice gas (GMLG) model.
We calculate the surface shape and flow profiles for a monodisperse system
and we introduce a novel generalization of the GMLG model for a binary
mixture of particles of different friction properties where, for the first
time, 
we measure the segregation process on the surface. 
The results are in good agreement with a recent theory, and we explain the
observed small deviations by the nonuniform velocity profile.

\end{abstract}


\bigskip

\begin{multicols}{2}

Mixtures of grains tend to separate as a response to virtually any type
of external perturbation~\cite{Herrmann}, an effect that can be a major
problem in some practical situations, and useful in others.  Indeed,
understanding the underlying processes of segregation phenomena in
granular mixtures is an intriguing problem of interest to scientists
from a wide range of disciplines.

Segregation occurs when a mixture is poured onto a horizontal base and a
pile builds up. When the grains of the mixture differ in size, the large
grains tend to gather at the bottom of the pile \cite{Brown}, while when
the grains have different shapes, the more faceted grains are found
preferentially near the top. An even more surprising effect can be
observed when a mixture of small rounded and large faceted grains is
poured between two parallel plates: Stratification is observed---i.e.,
the grains organize spontaneously into stripes~\cite{Makse}.

Particle size distribution crucially influences segregation. However, 
the particles generally differ not only in size but also in other properties, such as frictional ones. Frictional effects play a relevant role in band formation in 3-D rotating drums, in segregation in thin 
rotating drums, and also segregation and stratification in 2-D silos. Thus
the study of the particular case where segregation is caused solely by 
differences in friction coefficients tells the contribution of friction
to these segregation processes.

The description of segregation in the case of a bidisperse
pile is nontrivial.  A theory was proposed for the two-dimensional case
\cite{Boutreux}, which is a generalization of a method developed
\cite{Bouchaud} for avalanches in piles of monodisperse particles that
treats the static bulk and the fluidized surface (the rolling phase)
separately, and a set of continuum equations describes the dynamics of
the flowing region and the interactions between the two phases.
Solutions have been found for the steady state filling of a 2D silo for
the case of monodisperse particles. The theory provides also predictions
for the surface and segregation profiles in a binary mixture consisting
of grains with different friction properties but the same size
\cite{Boutreux}. 

It is important to test these theoretical predictions for two reasons.
First, the complete description requires a number of constitutive
relations between the relevant variables of the problem; these are
usually unknown and are assumed to have analytically treatable
functional forms, which should be verified.  Second, in order to derive
closed formulae for the profiles, several assumptions are made, so the
limits of these approximations are of interest.  Hence we have carried
out a program of computer simulations using the granular media lattice
gas (GMLG) model, which has been used for several granular systems, such
as pipe flows, shaken boxes, and static
piles~\cite{KK,Herrmann1,Herrmann2}. An advantage of the simulation
approach is that quantities can be measured that are inaccessible in
laboratory experiments.

In the GMLG model\cite{KK}, one generalizes a fully discrete
hydrodynamic algorithm~\cite{Frisch-lgpardif} in order to include energy
dissipation through particle collisions and friction.  The
indistinguishable point-like particles are either at rest or else they
travel with  unit momentum along the bonds of a triangular lattice.  The
particles are scattered at the lattice nodes at integer times and then
they are transferred to the nearest-neighbor sites in parallel.   

\begin{figure}
\centerline{\psfig{figure=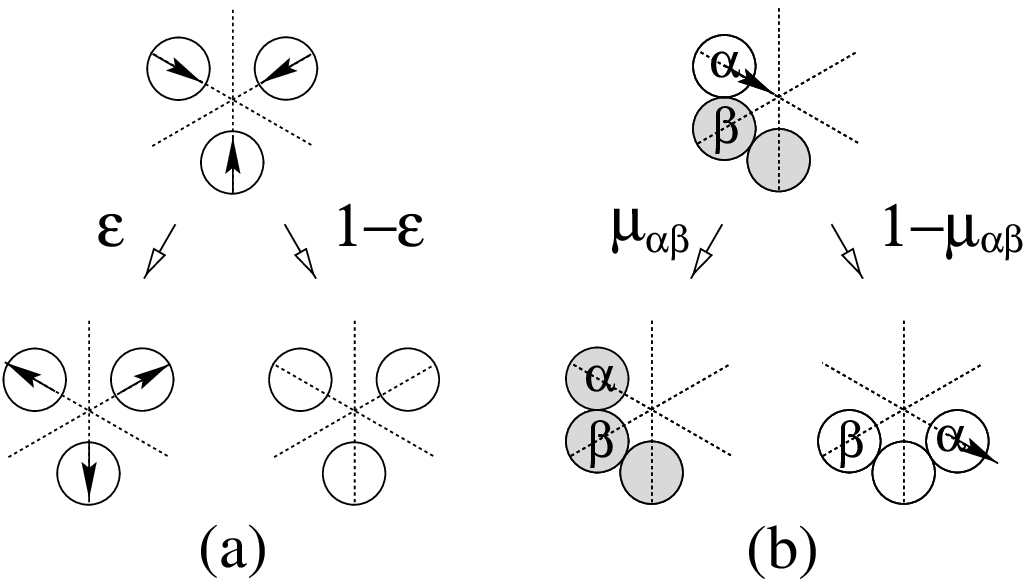}}
\caption{}
\label{colfric}
\end{figure}

We adapt the GMLG algorithm to the case of two types of particles with
different friction properties, which we call {\it up\/} and {\it down\/}
particles~\cite{Boutreux}.  Using probability variables, we introduce
material parameters in a stochastic way.  The restitution coefficient is
described by a parameter $\varepsilon$, which is the probability that
energy is conserved in a collision (an example of the application of
this rule is shown in Fig.~\ref{colfric}a). Momentum is conserved if the
particles in a collision are not connected, even in an indirect way, to
the wall.  

The compact static part of the pile behaves like a solid with a large
mass, where friction effects are taken into account.  When moving
particles interact with the bulk, their momentum can be transferred
through the force chains to the walls of the vessel.  For two types of
grains, we define four different friction coefficients
$\mu_{\alpha\beta},$ $(\alpha,\beta\in\uparrow,\downarrow)$, giving the
probability of a moving particle of type $\alpha$ to stop when arriving
at a bulk site containing a particle of type $\beta$
(Fig.~\ref{colfric}b).  Bulk particles are rest particles that are
supported by another bulk particle or by the vessel. The particles have
equal size, their distinction being introduced through different
friction coefficients~\cite{Boutreux}.

\begin{figure}
\centerline{\psfig{figure=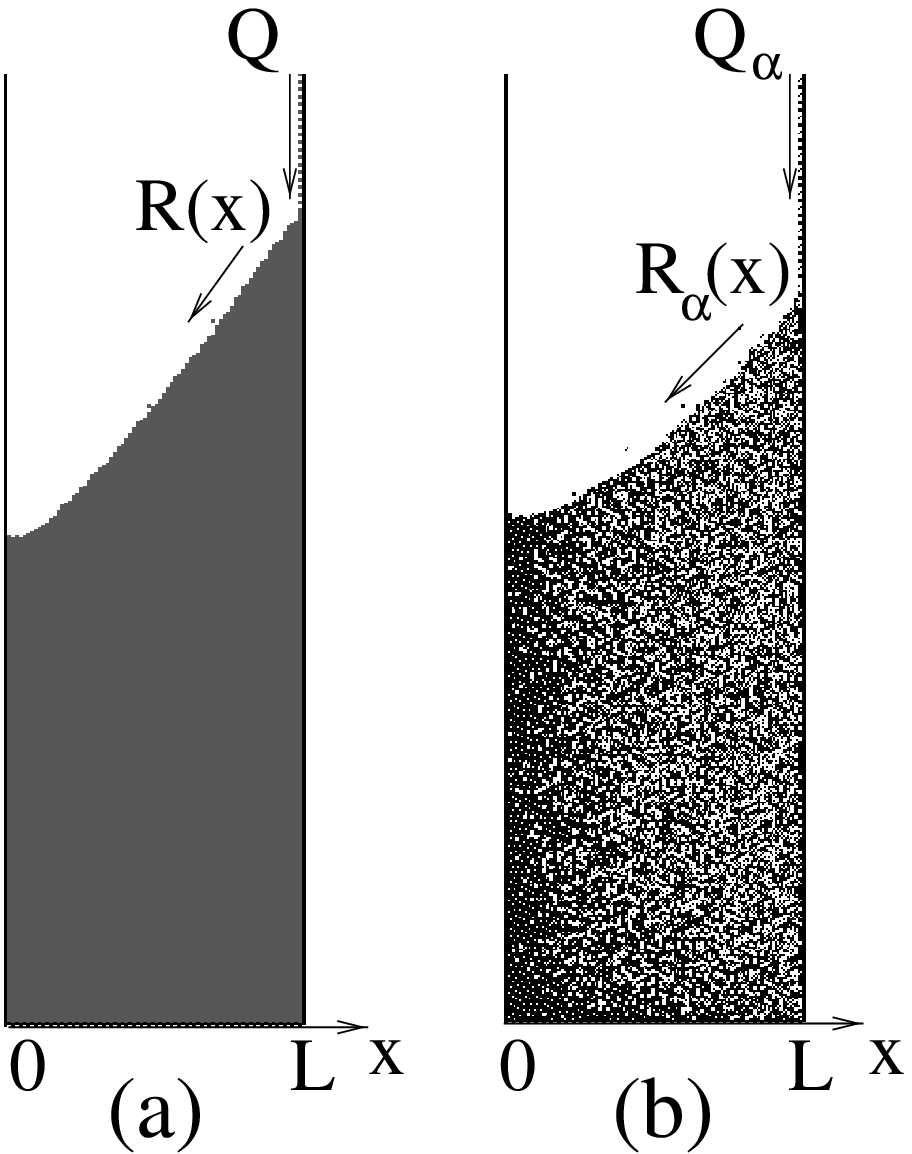}}
\caption{}
\label{silo}
\end{figure}

We study the steady state filling process of a two-dimensional ``silo'',
i.e., a long rectangular box of lateral size $L$. The silo is filled
with a steady flux of particles $Q$ next to the right wall. Two typical
snapshots are shown in Fig.~\ref{silo} for a monodisperse system of
particles and for a binary mixture.  
The theory focuses on the limit of very slow, but still continuous
filling.  In general, the steady state slope depends on the incoming
particle flux but this dependence vanishes at low rates. We 
find that for $Q \leq 0.5 {{\rm particle \over update~step}}$, the slope
will indeed be independent of the filling rate, while we still observe
smooth and non-intermittent growth. 
For even smaller incoming flux ($Q < 0.1$ in the above units) the pile grows intermittently. This avalanche regime was studied in Ref.~\cite{KK} in detail.

In steady filling of a monodisperse species, theory predicts that the
thickness of the rolling particle layer, $R(x)$, decreases 
linearly down the slope\cite{Boutreux}, and that the local slope $\theta(x)$ is close to the
angle of repose $\theta_{rep}$ everywhere but near the bottom of the pile
(see also Ref.~\cite{Herrmann2} for the case of a sandpile in an open
cell),
\begin{eqnarray}
R(x) = {Q\over v L} x,\qquad\qquad
\displaystyle{\theta(x) = \theta_{rep} - {v\over\gamma x}}.
\label{mono}
\end{eqnarray}
Here $\gamma$ describes the rate of exchange between the bulk and the
rolling phase, and $v/\gamma$ is of the order of the grain size.
We assume the flow velocity $v$ to be constant in space and time.

Next we test these predictions by calculating $R(x)$, defined in the
simulation by the average number of rolling particles at each $x$, and
the height of the pile, $\partial h(x)/\partial x\equiv \tan \theta(x)$.
We take data when the steady state is reached, and we repeat the measurements
at time intervals during which the pile grows about two lattice sites in
height. We average the data over an ensemble of $10-100$ systems.  The
number of particles at the end of each run is typically of the order of
$10^5$.

\begin{figure}
\centerline{\psfig{figure=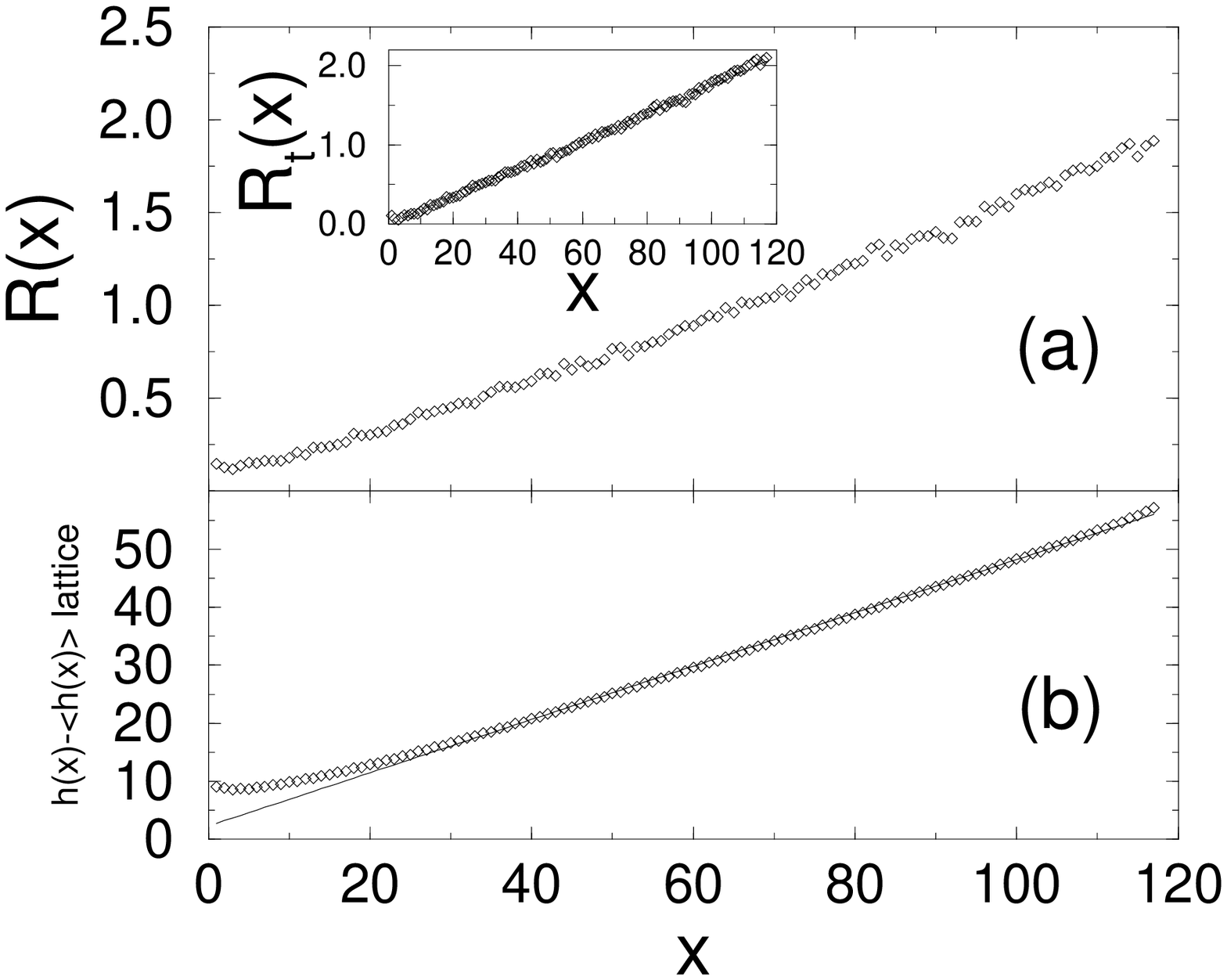}}
\caption{}
\label{pureprof}
\end{figure}

Figure \ref{pureprof} shows the simulated profiles for the monodisperse
case. Figure \ref{pureprof}a illustrates the fact that [see
Eq.~(\ref{mono})] 
\begin{equation}
\label{e.2}
R(x)\propto x
\end{equation}
with reasonable precision. The slight deviation from linearity can be
understood due to corrections to the velocity profile. Equation
(\ref{e.2}) comes from a conservation of grains argument, assuming
that the velocity of the rolling grains $v$ is constant along the slope.
However, at the bottom, where the slope is less steep we expect
the particles to slow down and this is also observed in the simulations. 
If we take this
into account theoretically by introducing a first-order correction to
the velocity, $ v(\theta) = v_0 + \lambda (\theta-\theta_{rep})$, we arrive
at an implicit formula for $R(x)$
(solutions for $\lambda=0$ were derived in Ref.\cite{Boutreux}):

\begin{equation}
R_t(x)\equiv 
R(x) + {{\lambda Q}\over{v_0 L\gamma}} \ln\left ( R(x) {v_0\over Q}\right) = 
{Q\over{v_0 L}} x.
\label{rtrans}
\end{equation}
By transforming our data with this solution, we obtain an excellent fit
(Fig.~\ref{pureprof}a inset).

From the height profile (Fig. \ref{pureprof}), 
we see that the slope at the upper part is
almost constant, and a region can be found at the bottom where the
surface flattens out. The singularity at $x=0$ [Eq. (\ref{mono})] cannot
be seen, since both in real systems and in simulations there is a cutoff
due to the grain size, but the profile does bend up (for small
$\varepsilon$) near $x=0$. This effect has also been observed
experimentally \cite{Herrmann3}. The singularity can be avoided by
re-defining the interaction term between the rolling and static grains.

If the silo is filled with a mixture of particles (Fig. \ref{silo}b),
the growth process is considerably more complicated. In general, instead
of one single angle of repose there are two continuous sets of
generalized angles of repose for the two species \cite{Makse} -- denoted by
$\theta_\uparrow$ and $\theta_\downarrow$ -- since the local critical
angles depend also on the volume fraction of the species in the bulk.  The
curves are characterized by four variables $\theta_{\alpha\beta}$, which
are the critical angles for particles of type $\alpha$ rolling on a pure
static phase consisting of grain type $\beta$. As in Ref.  \cite{Boutreux},
we assume that both curves are constant. Then the number of critical angles
reduces to two constants 
$\theta_\uparrow=\theta_{\uparrow\uparrow}=\theta_{\uparrow\downarrow}$ and
$\theta_\downarrow=\theta_{\downarrow\downarrow}=\theta_{\downarrow\uparrow}$
(or equivalently, $\mu_\uparrow$ and $\mu_\downarrow$) \cite{Makse}.

The rolling phase in the steady state is described by the number of
rolling particles for each species $R_\alpha (x)$ (with
$\alpha\in\uparrow,\downarrow$), and the static phase by the bulk volume
fractions $\Phi_\uparrow (x)$, and $\Phi_\downarrow (x)$, where
$\Phi_\uparrow + \Phi_\downarrow=1$. We distinguish two regions, where
different analytic results have been obtained \cite{Boutreux}. The {\it
outer region} includes almost the entire pile surface except for a
narrow zone, the {\it inner region}, close to the bottom of the pile. We
will focus on the flow properties in the outer region. The profile $R(x)
\equiv R_\uparrow(x) + R_\downarrow(x)$ is given by Eq.~(1) while the
profiles of the components can be expressed in terms of an exponent $r$:

\begin{eqnarray}
\label{rupx}
R_\uparrow(x)={R(x)\over{1+{Q_\downarrow\over Q_\uparrow} \bigl({L\over x} \bigr)^r}}, \\
\label{rdnx}
R_\downarrow(x)={R(x)\over{1+{Q_\uparrow\over Q_\downarrow}
\bigl({x\over L} \bigr)^r}}.
\end{eqnarray}
The exponent $r$ plays also a role in the determination of the bulk
volume fractions $\Phi_\uparrow$ and $\Phi_\downarrow$:
\begin{eqnarray}
\label{pupx}
\Phi_\uparrow(x) = \biggl(1+r {R_\downarrow(x)\over R(x)} \biggr) {R_\uparrow(x)\over R(x)},\\
\label{pdnx}
\Phi_\downarrow(x) = \biggl(1-r {R_\uparrow(x)\over R(x)} \biggr){R_\downarrow(x)\over R(x)}
\end{eqnarray}
Here we assume a homogeneous rolling phase with constant velocity
$v\equiv v_\uparrow = v_\downarrow$, and $Q_\uparrow$ and $Q_\downarrow$
are the fluxes of each species.  The exponent $r$ depends on the
structure of the collision matrix describing the interaction between the
bulk and rolling phase.

To analyze the simulation data, we calculate the exponent $r$ from the
measured $R(x)$ and $R_\alpha (x)$ profiles for all $x$.  The most
significant test of the theory is the existence of $r$; if $r$ is
well-defined, the volume fraction profiles can be calculated and
compared to the measured ones.

\begin{figure}
\centerline{\psfig{figure=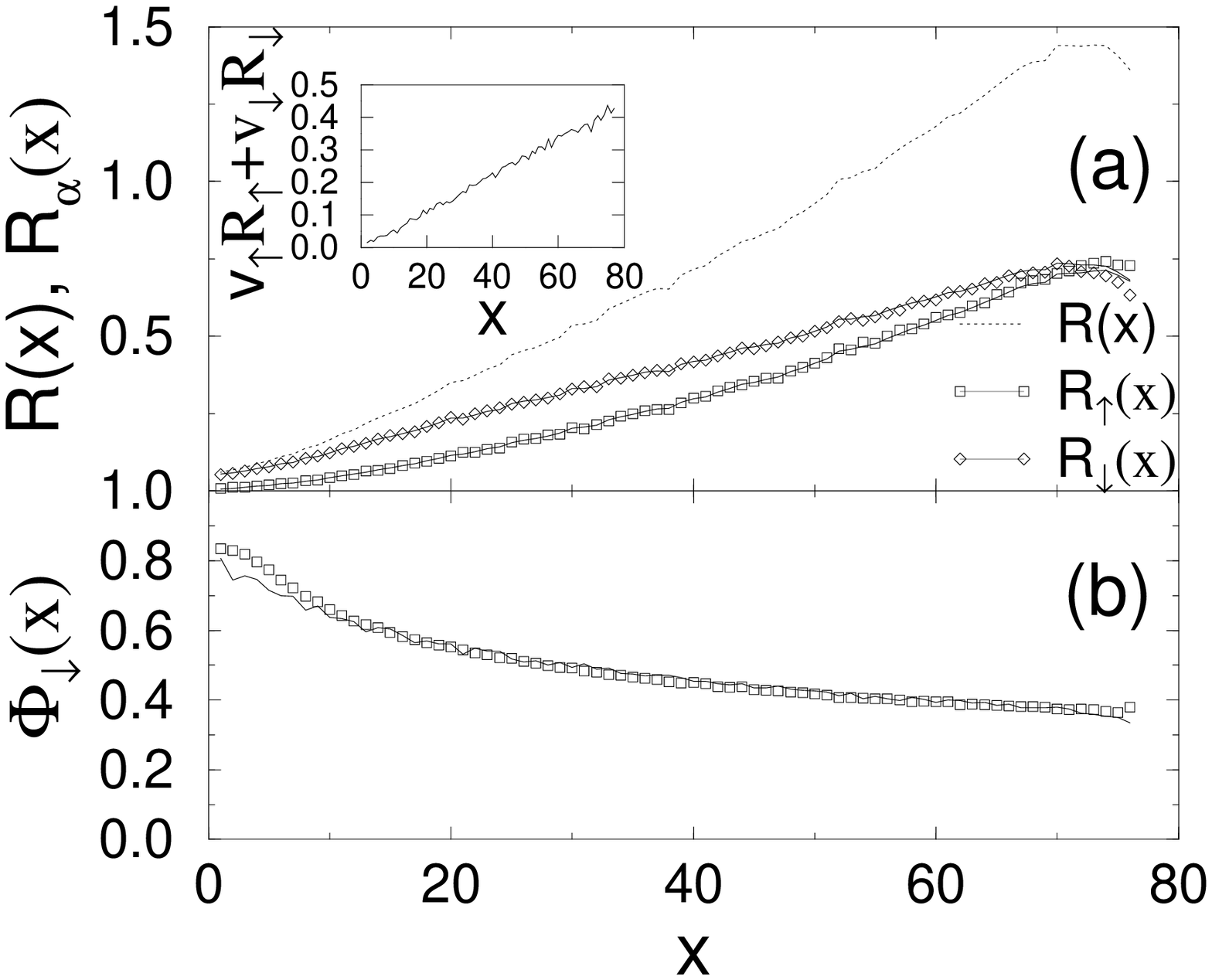}}
\caption{}
\label{prof}
\end{figure}

Figure \ref{silo}b shows a snapshot of the simulation for a mixture with
$\mu_\uparrow = 0.25$ ($\theta_\uparrow=57^{\rm o}$) and $\mu_\downarrow
= 0.152$ ($\theta_\downarrow=48^{\rm o}$). We see the segregation of the
mixture with the more sticky species ($\uparrow$) at the top of the
pile. Figure
\ref{prof}a shows typical moving particle profiles. It is apparent that
there is a slight higher order deviation from linearity in the case of
$R(x)$, as opposed to the prediction of Eq. (\ref{e.2}). The discrepancy
is small, but it is significant enough that the theoretical profiles
based on a linear approximation of the curve do not fit the simulation
results. However, if we use the {\it measured} $R(x)$ for calculating
the rolling particle profiles, Eqs. (\ref{rupx}) and (\ref{rdnx}) hold
to a better approximation (this bias will be discussed below).

\begin{figure}
\centerline{\psfig{figure=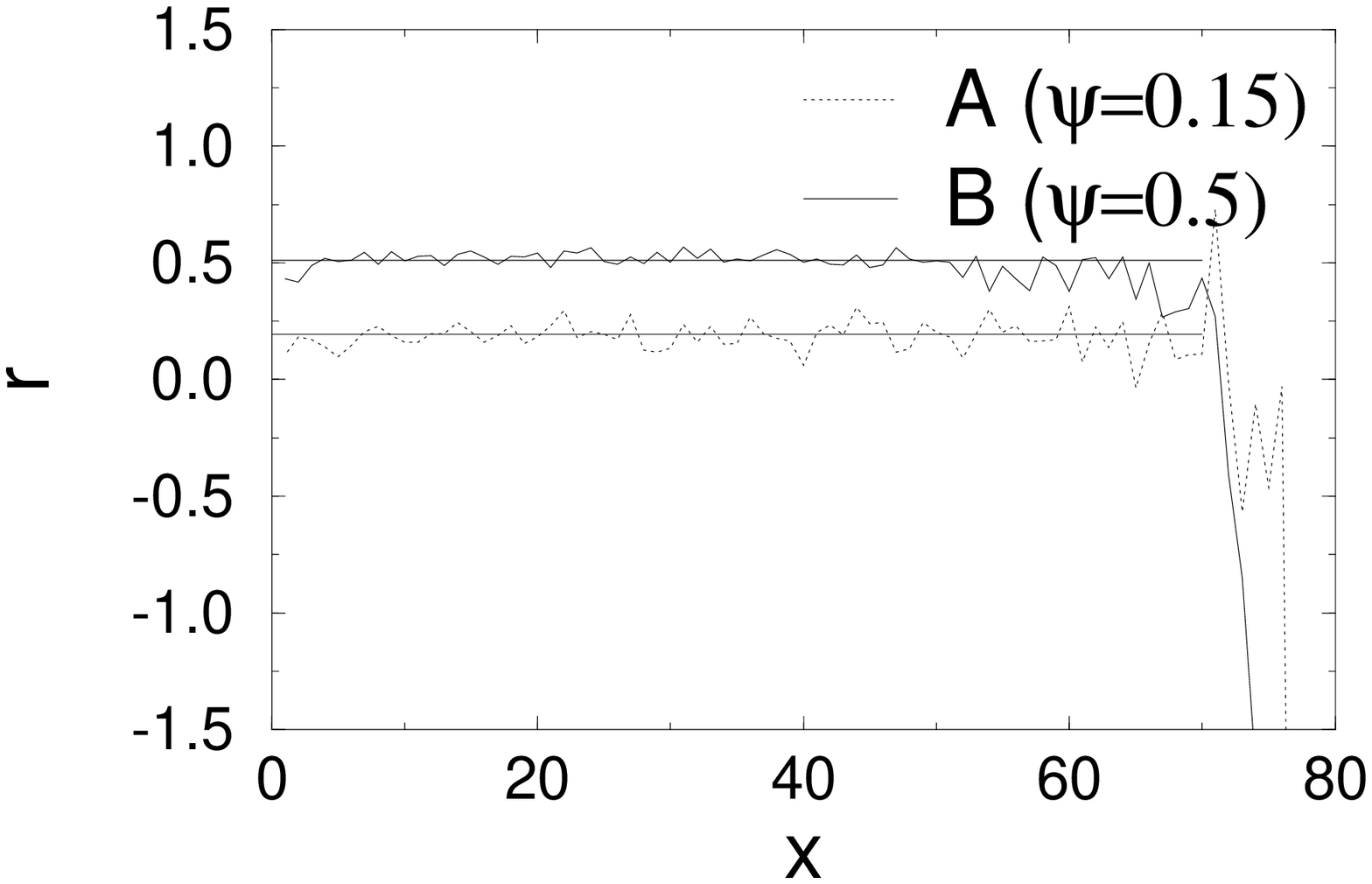}}
\caption{}
\label{rexp}
\end{figure}

Most crucial is to verify Eqs. (\ref{rupx}) and (\ref{rdnx}) by
calculating $r$. We present results for two sets of critical angles $A$
and $B$, with $\psi_A \approx 0.15$ and $\psi_B\approx 0.5$, where
$\psi\equiv\theta_\uparrow-\theta_\downarrow$.  Figure \ref{rexp}
demonstrates that the exponent is well defined in both cases except for
a region at the top of the pile. The measured exponents are $r_A = 0.19
\pm 0.02$ and $r_B = 0.51
\pm 0.03$. This result is reassuring since $r$ is expected to be of the order
of $\psi$
~\cite{Boutreux}.
The exponent slightly depends on the
$Q_\downarrow/Q_\uparrow$ ratio, but is independent of the total flux
provided $Q$ is sufficiently small.

With the help of the exponent we can obtain the volume fraction profiles
based on Eqs. (\ref{pupx}) and (\ref{pdnx}), and compare them to the
measured ones (Fig. \ref{prof}b). We find good agreement, except for a
slight deviation in the inner region. We find stronger segregation for
larger $\psi$ (or $r$); for the small angle difference $\psi_A$ the
segregation of the species is less pronounced.

Although the numerical results fit the continuum theory, we observe some
deviations. At the top of the pile, we see a discrepancy both at the
rolling particle profiles and when calculating the $r$ exponent. Here
the dynamics is significantly different from what is considered in the
continuum model: moving particles tend to be in free flight after
collisions with the pile surface.

Relation (\ref{e.2}) is not satisfied rigorously, as mentioned above.
The reason is that, similar to the monodisperse case, the $x$ component
of the velocity is not uniform. For a weighted sum of the rolling
particle densities, however, linearity should hold 
to a better approximation:
$v_\uparrow(x) R_\uparrow(x) +
v_\downarrow(x) R_\downarrow(x) \propto x$. 
We show a justification of this ansatz in Fig. \ref{prof}a, by using 
the measured $v_\uparrow(x)$ and $v_\downarrow(x)$ functions. 
The functional forms of these profiles can be approximated by 
$v_0-{a\over x}$, where $v_0=0.9 \pm 0.01 {\rm lattice~unit \over update~step} $ and $a=1.6 \pm 0.2 {\rm lattice~unit^2 \over update~step}$ are fitting parameters. The functional form of the velocities is consistent with the correction we found for the monodisperse system $v_\alpha(x) = v_0 + \lambda 
(\theta - \theta_\alpha)$, $\alpha\in\uparrow,\downarrow$, assuming that 
the angle of the pile behaves approximately as Eq. (\ref{mono}).

In general, we expect deviations from the theoretical predictions as an
increasing number of particles are in free flight above the pile. If
many particles tend to get detached from the surface due to elastic
collisions such a contribution should also be incorporated into the theory.
In the numerical results presented
above, both particle-particle and particle-wall collisions are almost
perfectly inelastic (the coefficient of restitution is around $0.25$).
Test runs 
for a more elastic medium show that the exponent
describing the $R_\alpha(x)$ profiles is no longer constant as a
function of $x$ indicating the limits of the theory.


In summary, we have simulated the filling process of a two-dimensional
silo, and our results for inelastic particles compare well with the
predictions of recent theories of surface flow of granular mixtures.  We
find small corrections that are accounted for by a modified set of
equations. Thus, we have verified the main assumptions of the theory,
pointed out the limits of the approximations involved, suggested
improvements, and found good agreement between the simulation results
and the improved theory. 
Future work is needed to test further the limitations of the theory
and to generalize the model to more complex situations like
polydispersity in the particle size distribution.

This work was supported by OTKA (T016568, T024004), MAKA (93b-352),  the
Heinrich-Hertz Stiftung, and the NSF.

\end{multicols}

\newpage

FIG.\ref{colfric} (a) Illustration of the implementation of the
restitution coefficient through a triple collision. If energy is
conserved---with probability $\varepsilon$---the particles are
scattered. If the collision is dissipative---with probability
($1-\varepsilon$)---the particles stop.  (b) An example of the application
of the friction rule, when a moving particle of type $\alpha$ (white)
arrives at a bulk site with a particle of type $\beta$ (shadow). With
probability $\mu_{\alpha\beta}$, the particle of type $\alpha$ loses
its energy and becomes part of the bulk, while with probability
$1-\mu_{\alpha\beta}$ its momentum is conserved and the node is not
considered to belong  any longer to the static phase.

FIG.\ref{silo} Two simulation snapshots for the steady filling of a
silo, which is being filled with (a) uniform particles (b) a mixture of
two different types of particles. Note that each pixel represents a
lattice node that can be occupied by up to six particles. In (b) black
and white dots denote lattice nodes where the $\downarrow$ and
$\uparrow$ particles are in majority, respectively ($\theta_\uparrow >
\theta_\downarrow$). The more sticky $\uparrow$ particles are found
preferentially at the top.

FIG.\ref{pureprof} Profiles of the growing pile in the steady state of
the monodisperse system. (a) The average number of rolling particles,
$R(x)$. The inset shows that $R_t(x)$ [see Eq. (\protect\ref{rtrans})]
is indeed proportional to $x$ to a good approximation, with fitting
parameter $\lambda=0.1
{\rm lattice~unit \over update~step}$.
(b) The height of the surface, $h(x)$. At each
measurement the mean height is subtracted, since the pile is constantly
growing.
On Figs.\ref{pureprof}-\ref{rexp} quantities are given in natural units of the simulation method: Lengths are measured in lattice constants, times in update steps. 

FIG.\ref{prof} Profiles in case of a granular mixture for $\psi_B
\approx 0.5$.  (a) Rolling particles. The continuous lines are fits
calculated using Eqs.  (\protect\ref{rupx}) and (\protect\ref{rdnx}).
The inset shows that $v_\uparrow R_\uparrow + v_\downarrow R_\downarrow
\propto x $.  (b) Volume fraction of the $\downarrow$ particles in the
static bulk. (For the $\uparrow$ particles $\Phi_\uparrow = 1 -
\Phi_\downarrow$.) The continuous line shows the calculated profile
according to Eq. (\protect\ref{pdnx}) using the measured $r$ exponent.

FIG.\ref{rexp} The value of the exponent $r$ calculated at each
site $x$ for $\psi_A=0.15$ and $\psi_B=0.5$; $r$ is well-defined
except for the uppermost part of the pile.

\end{document}